\documentclass[a4paper]{article}
\newcommand{\eg}{e.\,g., }
\newcommand{\ie}{i.\,e., }
\usepackage{INTERSPEECH2022}
\usepackage{hyperref}
\usepackage[backend=bibtex,citestyle=numeric-comp,bibstyle=ieee,sorting=nty,defernumbers=true,giveninits=true,doi=false,isbn=false,url=false,eprint=false,minbibnames=3,maxbibnames=3]{biblatex}
\usepackage[usestackEOL]{stackengine}
\usepackage{caption}
\title{SVTS: Scalable Video-to-Speech Synthesis}
\name{Rodrigo Mira$^1$, Alexandros Haliassos$^1$, Stavros Petridis$^{1}$, Björn W.\ Schuller$^{1,2}$, Maja Pantic$^{1}$}
\address{
  $^1$Imperial College London, UK\\
  $^2$Chair of Embedded Intelligence for Health Care and Wellbeing, University of Augsburg, Germany}
\email{\{rs2517,ah2214,stavros.petridis04,bjoern.schuller,m.pantic\}@imperial.ac.uk}

\begin{document}

\maketitle

\begin{abstract}
Video-to-speech synthesis (also known as lip-to-speech) refers to the translation of silent lip movements into the corresponding audio. This task has received an increasing amount of attention due to its self-supervised nature (\ie can be trained without manual labelling) combined with the ever-growing collection of audio-visual data available online. Despite these strong motivations, contemporary video-to-speech works focus mainly on small- to medium-sized corpora with substantial constraints in both vocabulary and setting. In this work, we introduce a scalable video-to-speech framework consisting of two components: a video-to-spectrogram predictor and a pre-trained neural vocoder, which converts the mel-frequency spectrograms into waveform audio. We achieve state-of-the art results for GRID and considerably outperform previous approaches on LRW. More importantly, by focusing on spectrogram prediction using a simple feedforward model, we can efficiently and effectively scale our method to very large and unconstrained datasets: To the best of our knowledge, we are the first to show intelligible results on the challenging LRS3 dataset.
\end{abstract}

\section{Introduction}
Lipreading, also known as visual speech recognition (VSR), is defined as the prediction of text transcriptions from silent video of lip movements. The advent of deep learning has enabled practitioners to shift from using only very constrained datasets \cite{DBLP:journals/corr/AssaelSWF16} to training models for lipreading in the wild \cite{DBLP:journals/corr/abs-2202-13084}. The progress in lipreading as well as text-to-speech (TTS) \cite{DBLP:conf/icassp/ShenPWSJYCZWRSA18} has drawn attention to the idea of predicting speech from silent video directly. This task, known as video-to-speech synthesis, has many impactful applications, such as generating clean speech when videoconferencing under noisy conditions, and helping people suffering from aphonia, who are unable to produce voiced speech. Although video-to-speech can be achieved through a combination of lipreading and text-to-speech, directly predicting speech obviates the need for labels (text transcriptions), meaning that it can be trained on raw video only.

To the best of our knowledge, the first work to train a neural network for video-to-speech synthesis was \cite{DBLP:conf/interspeech/CornuM15}, which predicts the audio clip's spectral envelope from a set of visual features extracted from video. It uses a stack of fully connected layers and feeds this envelope into a vocoder to produce voiced speech. This work was later extended in \cite{DBLP:journals/taslp/CornuM17}, achieving substantially more intelligible results for a single-speaker subset of GRID \cite{grid}. Following this, \cite{DBLP:conf/iccvw/EphratHP17} (an extension of another early video-to-speech approach \cite{DBLP:conf/icassp/EphratP17}) was the first to train and evaluate on multiple speakers (in this case, a 4-speaker subset of GRID), achieving a major leap forward in the realism of its outputs. This method set two trends which are widely adopted in following works: predicting speech features directly from raw video, rather than from manually extracted visual features \cite{DBLP:conf/icassp/AkbariACM18,DBLP:conf/interspeech/Vougioukas0PP19,DBLP:conf/icassp/YadavSNH21,DBLP:conf/cvpr/PrajwalMNJ20,DBLP:conf/interspeech/MichelsantiSHGT20,DBLP:conf/eusipco/OneataSC21,DBLP:journals/corr/abs-2107-12003,kim2021lip,DBLP:journals/taslp/HongKPR21}, and using mel-frequency spectrograms as an intermediate representation \cite{DBLP:conf/icassp/AkbariACM18,DBLP:conf/icassp/YadavSNH21,DBLP:conf/cvpr/PrajwalMNJ20,DBLP:conf/eusipco/OneataSC21}, which are then converted into raw waveform using the Griffin-Lim algorithm \cite{1164317}.
Notable exceptions include \cite{DBLP:conf/interspeech/Vougioukas0PP19}, which proposes an end-to-end video-to-waveform generative adversarial network (GAN) capable of producing intelligible speech from raw video without the need for a separate spectrogram-to-waveform system, and \cite{DBLP:conf/interspeech/MichelsantiSHGT20}, which uses a traditional vocoder to synthesize speech, rather than a spectrogram-based approach. 

Remarkably, most recent works focus on corpora with small pools of speakers, constrained vocabularies, and video recorded in studio conditions (\eg 4-Speaker GRID and 3-Lipspeaker TCD-TIMIT \cite{7050271}) \cite{DBLP:conf/icassp/AkbariACM18,DBLP:conf/interspeech/Vougioukas0PP19,DBLP:conf/icassp/YadavSNH21,DBLP:conf/interspeech/MichelsantiSHGT20,DBLP:conf/eusipco/OneataSC21,DBLP:journals/corr/abs-2107-12003,DBLP:journals/taslp/HongKPR21}, achieving improvements in performance via the use of intricate loss ensembles \cite{9760273,DBLP:journals/corr/abs-2107-12003,kim2021lip} and complex architectures \cite{DBLP:conf/cvpr/PrajwalMNJ20,DBLP:conf/icassp/YadavSNH21,DBLP:journals/corr/abs-2107-12003,DBLP:journals/taslp/HongKPR21}. While these developments are meaningful within ideal conditions, they fail to leverage the massive amount of audio-visual data available publicly, and propose training procedures which do not easily scale to very large datasets \cite{9760273,kim2021lip}. In this work, we aim to address these issues by proposing a simple video-to-speech system which efficiently scales with more data. It consists of a video-to-spectrogram predictor followed by a spectrogram-to-waveform synthesizer. The former is a ResNet18+conformer network \cite{DBLP:conf/cvpr/HeZRS16,DBLP:conf/interspeech/GulatiQCPZYHWZW20}, which becomes deeper and wider for larger datasets and is trained using a combination of two established comparative losses. The latter is a pre-trained neural vocoder, which accurately synthesizes the corresponding audio waveform with a low computational overhead.

Our contributions are as follows: \textbf{(1)} We present a simple and effective video-to-speech approach that can easily scale to large and complex datasets. \textbf{(2)} We conduct a detailed ablation study demonstrating the differences between commonly-used spectrogram inversion methods, as well as validating our choice of loss functions. \textbf{(3)} We outperform previous approaches on most metrics on the small but popular GRID dataset and achieve state-of-the-art performance on the larger LRW dataset. \textbf{(4)} To the best of our knowledge, we are the first to present intelligible results on the challenging LRS3 \cite{Afouras18d} dataset, and show that scaling our model even further with a combination of LRS3 and VoxCeleb2 \cite{Chung18b} (containing more than 1,500 hours of data) yields significant improvements.

\section{Methodology}
\begin{figure}[t]
\centering 
\includegraphics[width=0.8\linewidth]{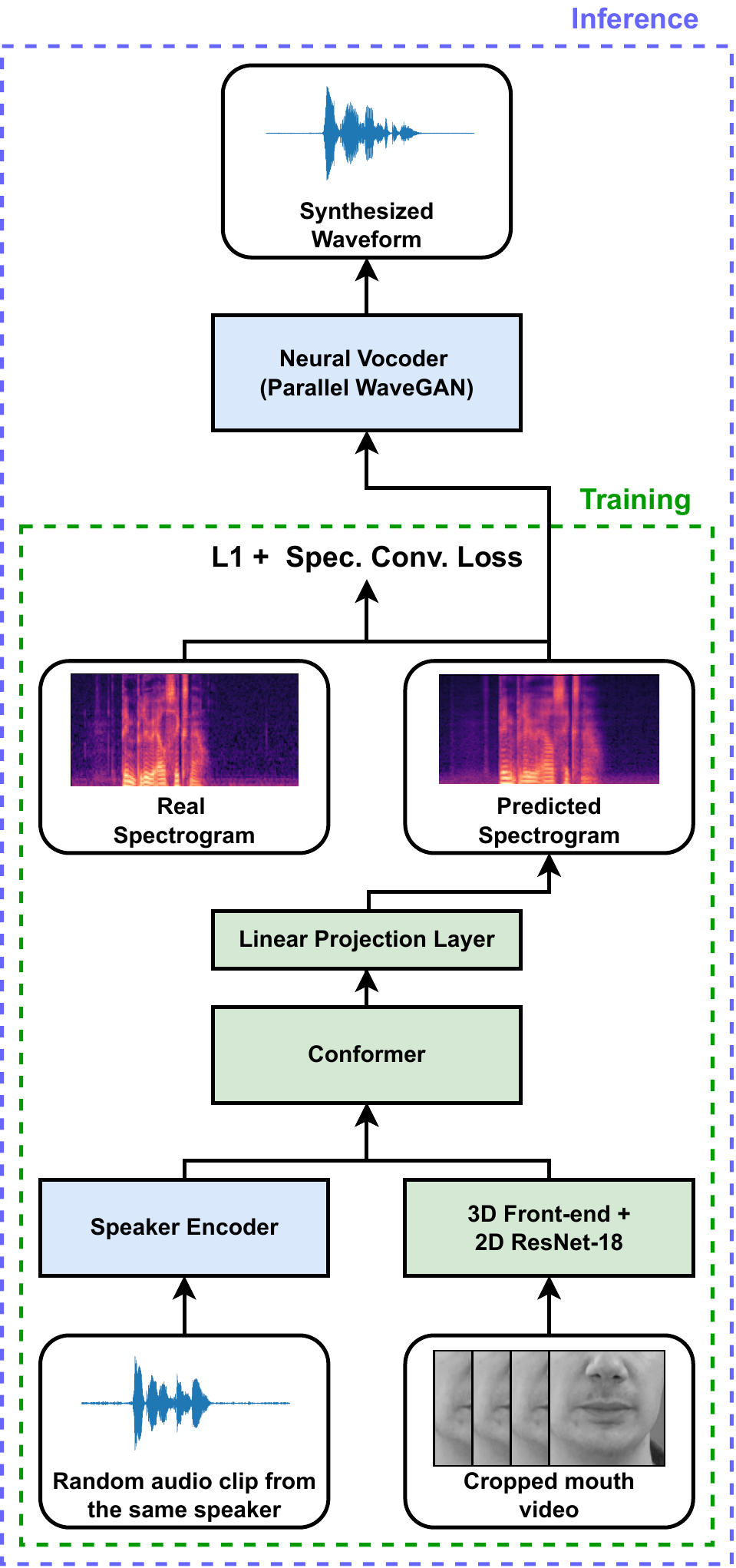}
\caption{Summary of our video-to-speech synthesis approach during training and inference. In this figure, the components pictured in blue are pre-trained and kept frozen, while the components pictured in green are trained from scratch.}
\label{architecture}
\vspace{-5mm}
\end{figure}
\subsection{Video-to-spectrogram model} \label{v2s_sec}
Our spectrogram predictor comprises two main components: (1) a visual encoder composed of a 3D convolutional stem followed by a standard 2D ResNet-18 \cite{DBLP:conf/cvpr/HeZRS16}, as in \cite{DBLP:journals/corr/abs-2202-13084}, and (2) a conformer \cite{DBLP:conf/interspeech/GulatiQCPZYHWZW20}, which receives the features from the visual encoder and aims to model the temporal correlations between them. The latter contains an initial linear layer, followed by a set of conformer blocks which vary in depth and width based on the model version, as shown in Table \ref{architecture_table}. Finally, each feature vector, corresponding to a video frame, is projected into a hidden size of $320$ using a linear projection layer, and reshaped into $4\times80$ spectrogram frames. The input video is sampled at 20 fps and the extracted spectrogram contains 80 frames per second. We train our predictor using a combination of the $L_1$ loss and the spectral convergence loss \cite{DBLP:conf/icassp/YamamotoSK20}.

As in multi-speaker text-to-speech systems \cite{DBLP:conf/nips/JiaZWWSRCNPLW18}, our video-to-speech model requires information about the speaker's voice characteristics, which cannot be derived accurately from silent video only. To this end, we use a pre-trained speaker encoder\footnote{\url{https://github.com/CorentinJ/Real-Time-Voice-Cloning}.} originally trained for speaker verification on a combination of VoxCeleb \cite{DBLP:conf/interspeech/NagraniCZ17}, VoxCeleb2 \cite{Chung18b}, and Librispeech \cite{DBLP:conf/icassp/PanayotovCPK15}. For each video clip, an embedding is extracted from a randomly selected audio clip from the same speaker and concatenated with the visual features extracted by the visual encoder, which are then fed into the conformer. Note that the speaker encoder is kept frozen during training.
\begin{table}[t]
  \caption{Summary of our proposed SVTS architectures. $^{*}$refers to the total number of parameters in the model (ResNet + conformer + projection layer)}.
  \centering
  \resizebox{\linewidth}{!}{
  \begin{tabular}{l|ccc}
    \toprule
    \textbf{Model}      & \textbf{SVTS-S} & \textbf{SVTS-M} & \textbf{SVTS-L}\\
    \midrule
    Num. parameters$^{*}$ (M)     & 27.3 & 43.1 & 87.6 \\
    Conformer blocks      & 6 & 12 & 12 \\
    Attention dim. & 256 & 256 & 512 \\
    Attention heads & 4 & 4 & 8 \\
    Conv. kernel size & 31 & 31 & 31 \\
    Feedforward dim. & 2048 & 2048 & 2048 \\ 
    
    \bottomrule
  \end{tabular}
  }
  \vspace{-5mm}
  \label{architecture_table}
\end{table}

\subsection{Spectrogram-to-waveform}
In order to generate waveform speech from the spectrograms, we opt for the use of a neural vocoder, specifically Parallel WaveGAN \cite{DBLP:conf/icassp/YamamotoSK20}. 
This WaveNet-based \cite{DBLP:conf/ssw/OordDZSVGKSK16} model is trained using a combination of comparative and adversarial losses. We employ a version pre-trained on LibriTTS \cite{DBLP:conf/interspeech/ZenDCZWJCW19} for 1 million iterations. Note that it is used only at inference time, allowing for a substantially simpler training procedure than related video-to-speech works, which train their own vocoder from scratch \cite{DBLP:journals/corr/abs-2107-12003,DBLP:journals/taslp/HongKPR21}. An overview of our approach is illustrated in Figure \ref{architecture}.

\section{Experimental setup}
\subsection{Datasets} \label{subsec:datasets}
The first corpus we experiment with is GRID, which has become an established benchmark in video-to-speech literature due to its small vocabulary, predictable structure, and clean recording conditions. GRID is composed of 1,000 unique sentences (with a small vocabulary of 51 words) uttered by 33 speakers; this amounts to roughly 27 hours of audio-visual speech. We experiment with two versions of the dataset: (1) a seen speaker version, originally proposed in \cite{9760273}, where the 33 speakers are present in the training, validation, and testing sets, and (2) an unseen speaker version, introduced in \cite{DBLP:conf/interspeech/Vougioukas0PP19}, where there is no overlap in the speakers between the sets.

The second corpus is LRW, which features around 150 hours of single-word utterances from hundreds of different speakers recorded `in the wild.' Although its 500-word vocabulary is not extensive, the filming conditions are significantly less controlled than GRID, with varying lighting, head poses, and background noise. As a result, LRW is considered more challenging than GRID and is substantially closer to a real-world scenario. Due to LRW's lack of speaker labels, it is not possible to select a random audio clip from the same speaker to produce the corresponding speaker embedding. Therefore, for this corpus we generate the speaker embeddings using the audio clip from the corresponding video, which is consistent with previous multi-speaker video-to-speech approaches on LRW \cite{DBLP:conf/cvpr/PrajwalMNJ20}.

To demonstrate our method's scalability to even larger and less constrained datasets, we run experiments on the 312-hour-long LRS3 dataset. It contains long sentences, a diverse vocabulary of more than 50,000 words, and thousands of speakers. As in GRID, we use two different versions of LRS3: seen speaker, where all speakers' utterances are split into training, validation and testing sets using a 80 -- 10 -- 10\,\% ratio, and unseen speaker, following the original split proposed in \cite{Afouras18d}. Finally, we experiment with combining the LRS3 training dataset with an English-only version \cite{DBLP:journals/corr/abs-2201-02184} of VoxCeleb2 (while keeping the same LRS3 validation and test sets to ease comparison), amounting to around 1,550 hours of footage. For both corpora, utterances exceeding 24 seconds are excluded from training due to hardware limitations.

\subsection{Data pre-processing and augmentation}
In order to produce the cropped mouth video, we first extract 68-point landmarks using RetinaFace\footnote{\url{https://github.com/biubug6/Pytorch_Retinaface}} \cite{DBLP:conf/cvpr/DengGVKZ20} and a pre-trained 2D-FAN\footnote{\url{https://github.com/1adrianb/face-alignment}} \cite{bulat2017far}. We average the landmarks across 12 frames through a sliding window to reduce motion jitter, and align each frame to the mean face. We then crop a $96\times96$ region centred around the mouth and convert the frames to grayscale. The audio is sampled at 24 kHz, and the log-mel spectrograms are extracted using 80 mel bands, frequency bins of size 2048, a hop size of 12.5 ms, a window length of 50 ms, and a Hann window.

During training, we apply random cropping of size $88\times88$, horizontal flipping with probability of 0.5, and random erasing with a probability of 0.5. The erased area is randomly sampled between 2 and 33\,\% of the full frame, with an aspect ratio ranging from 0.3 to 3.3. During testing, we perform center cropping of size $88\times88$. For our LRS3 experiments, we apply time-masking by randomly replacing each frame with the average pixel value in the video, since we find it aids generalization when training on long sentences. We apply one contiguous time-mask for each second of the utterance, and each mask's length is uniformly sampled from 0 to 0.4 seconds.
\subsection{Training details}
For our GRID and LRW experiments, we train our models using AdamW \cite{loshchilov2017decoupled} with a learning rate of $1\times10^{-3}$, $\beta_1 = 0.9$, $\beta_2 = 0.98$, and a weight decay of $1\times10^{-2}$. We warm up the learning rate for the first 10\,\% of iterations, and then decay it with a cosine schedule \cite{loshchilov2016sgdr}. For LRS3 seen speakers, we use a maximum learning rate of $7\times10^{-3}$, while for unseen speakers (including the combination with VoxCeleb2) we use $1\times10^{-3}$. We train for a total of 200, 150, 500, and 150 epochs for GRID, LRW, LRS3 seen speakers, and LRS3 unseen speakers, respectively. We save a checkpoint at the end of each epoch, and at the end of training select the one with the lowest validation loss.
\subsection{Evaluation metrics}
We measure the quality and accuracy of our generated samples via 4 objective metrics. The first is Perceptual Evaluation of Speech Quality (PESQ)\footnote{\url{https://github.com/ludlows/python-pesq}} \cite{DBLP:conf/icassp/RixBHH01}, which aims to measure the clarity and perceptual quality of the generated samples. We also use Short-Time Objective Intelligibility (STOI)\footnote{\url{https://github.com/mpariente/pystoi}} \cite{DBLP:journals/taslp/TaalHHJ11} and its extended version ESTOI to measure the intelligibility of our samples.

The final metric we apply is word error rate (WER), which has become a benchmark in video-to-speech after its introduction in \cite{DBLP:conf/interspeech/Vougioukas0PP19}. It is measured by applying a pre-trained speech recognition model to the generated samples, and comparing the predicted transcription with the ground truth. Hence, WER serves as an easily interpretable intelligibility metric for the generated samples. We propose to forego the use of manual text transcriptions, and use instead the transcription predicted from the corresponding real audio as the ground truth. This increases the interpretability of the reported numbers, as they are a direct measure of the difference in intelligibility between real and generated audio, and it also removes the requirement for labelled datasets in future work. For our GRID experiments, we use a model pre-trained on LRW, LRS2, and LRS3 \cite{DBLP:journals/corr/abs-2102-06657}, and fine-tuned on GRID (adopting the split from \cite{DBLP:journals/corr/AssaelSWF16}); it achieves a WER of 0.1\,\% on the real audio test set. For LRW, we use an ASR model trained only on LRW \cite{DBLP:conf/icassp/PetridisSMCTP18} with a WER of 1.68\,\%. 

Although these metrics are commonly referenced in video-to-speech works and are therefore useful for comparison, it is widely known that no objective speech metric correlates perfectly with human perception of quality and intelligibility \cite{DBLP:conf/interspeech/Vougioukas0PP19}. Therefore, we highly encourage readers to listen to the generated samples available on our project website\footnote{\url{https://sites.google.com/view/scalable-vts}} rather than rely solely on the reported metrics.
\section{Results} \label{results_section}
\subsection{Experiments}
\begin{table*}
  \caption{Summary of our results. Due to LRS3's complex vocabulary and long sentence structure, we are unable to find a speech recognition model that works accurately on our generated samples (\eg the word "teacher" is often mistaken for "teachers"), and therefore do not report WER for this dataset. $^{*}$reported using Google speech-to-text API.}
  \centering
  \resizebox{\linewidth}{!}{
  \begin{tabular}{cccc|cccc}
    \toprule
    \textbf{Method} & \textbf{Corpus} &  \Centerstack{\textbf{Speaker split} \\ \textbf{(seen/unseen)}} & \Centerstack{ \textbf{Training data} \\ \textbf{(hours)}} & \textbf{PESQ} & \textbf{STOI} & \textbf{ESTOI} & \textbf{WER (\%)}\\
    \midrule
    \midrule
    End-to-end GAN \cite{9760273} & GRID & seen & 24 & 1.70 & 0.667 & 0.466 & 4.60 \\
    VCA-GAN + Griffin-Lim \cite{kim2021lip} & GRID & seen & 20 & \textbf{1.97} & 0.695 & 0.505 & 5.13 \\
    SVTS-S & GRID & seen & 24 & \textbf{1.97} & \textbf{0.705}& \textbf{0.523} & \textbf{2.36} \\
    \midrule
    End-to-end GAN \cite{DBLP:conf/interspeech/Vougioukas0PP19} & GRID & unseen & 13 & 1.26 & 0.494 & 0.198 & 32.79 \\
    Conv. + GRU + WORLD vocoder \cite{DBLP:conf/interspeech/MichelsantiSHGT20} & GRID & unseen & 13 & 1.26 & 0.541 & 0.227 & 38.15 \\
    End-to-end GAN \cite{9760273} & GRID & unseen & 13 & 1.37 & 0.568 & 0.289 & \textbf{16.12}\\
    VCA-GAN + Griffin-Lim \cite{kim2021lip} & GRID & unseen & 13 & 1.39 & 0.570 & 0.282 & 24.57 \\
    Conv. + LSTM + WaveNet \cite{DBLP:journals/taslp/HongKPR21} & GRID & unseen & 13 & 1.33 & 0.531 & 0.271 & 26.17 \\
    SVTS-S & GRID & unseen & 13 & \textbf{1.40} & \textbf{0.588} & \textbf{0.318} & 17.85 \\
    \midrule
    Conv. + LSTM + Griffin-Lim \cite{DBLP:conf/cvpr/PrajwalMNJ20} & LRW & unseen & 157 & 1.20 & 0.543&  0.344 & 34.20$^{*}$ \\
    End-to-end GAN \cite{9760273} & LRW & unseen & 157 & 1.33 & 0.552 & 0.330 & 42.60 \\
    VCA-GAN + Griffin-Lim \cite{kim2021lip} & LRW & unseen & 157 & 1.34 & 0.565 & 0.364 & 37.07 \\
    SVTS-M & LRW & unseen & 157 & \textbf{1.49} & \textbf{0.649} & \textbf{0.483} & \textbf{13.40} \\
    \midrule
    SVTS-L & LRS3 & seen & 256 & \textbf{1.30} & \textbf{0.553} & \textbf{0.331} & - \\
    \midrule
    SVTS-L & LRS3 & unseen & 296 & 1.25 & 0.507 & 0.271 & - \\
    SVTS-L & \Centerstack{LRS3 +\\ VoxCeleb2} & unseen & 1556 & \textbf{1.26} & \textbf{0.530} & \textbf{0.313} & - \\
    
    \bottomrule
    \vspace{-5mm}
  \end{tabular}
  }
  \label{results_table}
\end{table*}

\begin{table}[t]
  \caption{Vocoder ablation on GRID (seen speakers). Speed is measured on an Nvidia RTX 2080 Ti. $^*$computed on CPU}.
  \centering
  \resizebox{\linewidth}{!}{
  \begin{tabular}{l|ccccc}
    \toprule
    \textbf{Metric}      & \textbf{PESQ} & \textbf{STOI} & \textbf{ESTOI} & \Centerstack{\textbf{WER} \\ \textbf{(\%)}} & \Centerstack{\textbf{Speed} \\ \textbf{(clips/sec.)}} \\
    \midrule
    Griffin-Lim$^*$ \cite{1164317}       & \textbf{2.00} & 0.696 & 0.513 & 2.41 & 1.2\\
    Multiband MelGAN \cite{DBLP:conf/slt/YangYLF0X21}   & 1.86 & 0.683 & 0.487 & 2.50 & \textbf{184.9}\\
    Style MelGAN \cite{DBLP:conf/icassp/MustafaPF21}       & 1.93 & 0.702 & 0.520 & 2.38 & 83.7\\
    Parallel WaveGAN \cite{DBLP:conf/icassp/YamamotoSK20}   & 1.97 & \textbf{0.705} & \textbf{0.523} & \textbf{2.36} & 54.7\\
    \bottomrule
  \end{tabular}
  }
  \label{vocoder_ablation}
  \vspace{-2mm}
\end{table}

\begin{table}[t]
  \caption{Loss ablation on GRID (seen speakers).}.
  \centering
  \resizebox{\linewidth}{!}{
  \begin{tabular}{l|cccc}
    \toprule
    \textbf{Metric}      & \textbf{PESQ} & \textbf{STOI} & \textbf{ESTOI} & \textbf{WER (\%)} \\
    \midrule
    w/o Spec. Conv.      & \textbf{1.97} & \textbf{0.705} & \textbf{0.523} & 2.90 \\
    w/o $L_1$              & 1.91 & 0.700 & 0.514 & 2.74 \\
    $L_1$+Spec. Conv.            & \textbf{1.97}& \textbf{0.705} & \textbf{0.523} & \textbf{2.36} \\
    \bottomrule
  \end{tabular}
  }
  \label{loss_ablation}
  \vspace{-5mm}
\end{table}
Our results are presented in Table \ref{results_table}. We begin by discussing our findings on the small-scale GRID dataset. For the seen speaker split, our SVTS-S model clearly outperforms our previous approach \cite{9760273}, as well as the more recent \cite{kim2021lip}, on STOI and ESTOI. It also achieves a significant improvement in WER. These metrics indicate that our samples are more intelligible than previous works. On unseen speakers, our model achieves a better PESQ, STOI, and ESTOI but is outperformed by our previous GAN-based work \cite{9760273} in WER. 
By perceptually evaluating the generated samples, we find that our seen speaker reconstruction is highly realistic and could be mistaken for real audio. On the other hand our unseen speaker samples sound considerably less noisy than previous works and capture the unseen speaker's voice with remarkable accuracy, thanks to our speaker embedding strategy.

On the more challenging and diverse LRW dataset, SVTS-M is superior to previous approaches on all metrics by a wide margin. We achieve a low WER of 13.4\,\%, indicating that our samples are consistently intelligible. Perceptually, we find that our samples sound substantially more realistic and accurate than previous approaches, including our GAN-based approach \cite{9760273}. This strong performance is a consequence of our SVTS architecture, which allows us to efficiently scale to this larger dataset.

Finally, we experiment with LRS3, which is undoubtedly the most challenging corpus we approach, as discussed in Section \ref{subsec:datasets}. On the seen speaker setting, we find that our model achieves reasonable PESQ, STOI and ESTOI performance, comparable to what had been reported by previous works on LRW. The unseen speaker protocol is naturally more challenging, and therefore does not achieve the same level of quality. Interestingly,  we find that results are greatly improved with the addition of the VoxCeleb2 data, as shown by the significant boost on all metrics. This empirically demonstrates our model's ability to improve its reconstructions by leveraging additional training data, even if its distribution is different from the testing set (which only contains samples from LRS3). It also suggests that we may have not yet reached a saturation point: There are likely still gains to be made in the future with even more data.

Perceptually, we find that the most intelligible samples are produced by our seen speaker model, closely followed by our model trained on LRS3+VoxCeleb2. Although there is room for improvement, we find that most syllables in the reconstructed speech are discernible, and each speaker's voice profile is reproduced with considerable accuracy, which is particularly impressive in the unseen speaker scenario.
\subsection{Ablations}
In order to motivate our use of Parallel WaveGAN (PWG) as our waveform synthesis model, we compare it in Table \ref{vocoder_ablation} with other recently proposed neural vocoders as well as the commonly used Griffin-Lim algorithm. All models, including our version of PWG, are pre-trained on LibriTTS and are publicly available\footnote{\url{https://github.com/kan-bayashi/ParallelWaveGAN}}. The Griffin-Lim synthesis is performed using the fast version of the algorithm\footnote{\url{https://librosa.org/doc/main/generated/librosa.griffinlim.html}} \cite{DBLP:conf/waspaa/PerraudinBS13}, and runs for 30 iterations. It can be observed that Parallel WaveGAN outperforms its peers Multiband Melgan \cite{DBLP:conf/slt/YangYLF0X21} and Style Melgan \cite{DBLP:conf/icassp/MustafaPF21} on all four evaluation metrics. Furthermore, through perceptual evaluation, we find that PWG produces substantially more realistic audio. Regarding Griffin-Lim, although it achieves a slightly higher PESQ score, we find that it consistently produces noisy speech with frequent artifacts. This highlights the limitations of PESQ as a metric, as it is often not sensitive to artifacts that are immediately noticeable to human listeners. Thanks to efficient GPU implementations, the vocoders are roughly $50\times$ faster than Griffin-Lim, with the fastest vocoder, Multiband Melgan, being able to process almost 200 GRID clips per second.

In Table \ref{loss_ablation}, we experiment with each of our loss functions separately and compare with the combined loss (baseline). We find that the baseline's performance is 
roughly similar to the individual losses on PESQ, STOI and ESTOI, but is clearly superior on WER. Interestingly, we find that our model achieves comparable performance with only an $L_1$ loss, which contrasts greatly with previous approaches' reliance on elaborate loss combinations \cite{9760273,DBLP:journals/corr/abs-2107-12003}.

\section{Conclusion}
In this paper, we propose SVTS, a scalable approach for video-to-speech synthesis. We present three architectures of varying sizes, which allow us to efficiently adapt our training procedure to datasets ranging from GRID (27 hours) to LRS3+VoxCeleb2 ($>$ 1,500 hours). We show that our method outperforms previous approaches on most metrics for two popular versions of GRID, and establishes a new state-of-the-art for LRW. Finally, we experiment with the large and unconstrained LRS3 corpus, achieving intelligible results, and combine it with VoxCeleb2 to further improve our performance, demonstrating our method's scalability. We hope our work will encourage a shift towards larger corpora, as this aligns with the current ubiquity of unlabelled audio-visual data.
\section{Acknowledgements}
We would like to thank Pingchuan Ma for sharing his speech recognition models. All training, testing, and ablation studies have
been conducted at Imperial College London.

\clearpage
\AtNextBibliography{\footnotesize}
\section{References}
\printbibliography[heading=none]
%{\footnotesize
%\bibliographystyle{IEEEtran}
%\bibliography{refs}
%}

\end{document}